\documentclass{article}

\usepackage{tikz}
\usepackage{spconf,amsmath,graphicx,booktabs,multirow,amsfonts,caption}
\usepackage{subcaption}

\usepackage{subfiles}
\usepackage{pgfplots} 
\usepackage{pgfplotstable} 
\pgfplotsset{compat=newest} 
\usetikzlibrary{plotmarks} 
\usetikzlibrary{colorbrewer} 
\usepackage{booktabs} 
\usepackage{etoolbox,siunitx} 

\robustify\bfseries
\robustify\itshape
\usepackage{paralist} 

\usepackage[inline]{enumitem}
\usepackage{flushend}

\definecolor{r1}{RGB}{228,26,28}
\definecolor{r2}{RGB}{55,126,184}
\definecolor{r5}{RGB}{20,175,20}
\definecolor{r20}{RGB}{152,78,163}

\usepackage{arydshln}
\usepackage[pagebackref=true,breaklinks=true,letterpaper=true,colorlinks,bookmarks=false]{hyperref}
\usepackage{xr}

\graphicspath{{figures/}}

\title{Tracked instance search}
\name{Andreu Girbau\thanks{The first author performed the work while at the NII, Tokyo}$^{\dagger}$ \qquad Ryota Hinami$^{\star}$ \qquad Shin'ichi Satoh$^{\star}$}

\address{$^{\dagger}$ Universitat Polit\`{e}cnica de Catalunya, Barcelona \\
    $^{\star}$ National Institute of Informatics, Tokyo}

\begin{document}
\ninept
\maketitle

\begin{abstract}

{ \it In this work we propose tracking as a generic addition to the instance search task. From video data perspective, much information that can be used is not taken into account in the traditional instance search approach. This work aims to provide insights on exploiting such existing information by means of tracking and the proper combination of the results, independently of the instance search system. We also present a study on the improvement of the system when using multiple independent instances (up to 4) of the same person. Experimental results show that our system improves substantially its performance when using tracking. Best configuration improves from $mAP={\bf 0.447}$ to $mAP={\bf 0.511}$ for a single example, and from $mAP={\bf 0.647}$ to $mAP={\bf 0.704}$ for multiple (4) given examples.}

\end{abstract}

\section{Introduction}
\documentclass[Tracked_INS.tex]{subfiles}

An important need in many situations involving video collections (archive video search/reuse, surveillance, law enforcement, protection of brand/logo use...) is to find images or video segments of a certain specific person, object, or place, given a visual example. To this purpose, instance search is defined as the problem of finding instances of the specific query in a set of images or videos given a visual example. 

Typically, an instance search system takes a query given as an image (or images) optionally region specified (rectangle or segment), and returns a ranked list of possible instances of that query from a dataset. A main problem is that, usually, only a single image is provided, and the results become solely dependent to the specific characteristics of that image (pose, illumination, viewpoint...). In other words, the system becomes very dependent towards the matching between the given visual example and the dataset.

For queries coming from videos, we propose to exploit the already existing video data. The objective is to provide more variance to the system, so that it is not so biased towards a single visual example. By applying tracking, we can collect many sample images of the target instance with sufficient variation which may result in better instance search performance. In figure \ref{fig:figure1} we show the expected mapping of the query expansion by using tracking to the feature space. 

In order to merge multiple ranked lists obtained by multiple sample images we propose to use a voting scheme. In this paper we study two possible voting schemes, one assuming dependence between the examples provided by tracking and the other assuming independence between them. We found that, by considering dependence between samples from tracking, our system achieves better performance than considering these samples as independent.

We developed a baseline to test the hypothesis on whether tracking helps or not in the instance search task. We used the TRECVID \cite{2016trecvidawad} instance search task for this purpose. From 2016, TRECVID INS task is based on retrieving a specific person in a specific location. This means that the correct person (i.e. the query) will be only tagged as good if he/she is in the specified location. In order to correctly evaluate our method we generated a person ground truth for this purpose based on the TRECVID INS dataset. Such ground truth will be made publicly available to help further research. 

Our contribution: we provide insights on the influence of tracking as a generic and automatic query expansion for video instance search, and propose a way to combine the results. Also we show the differences on performance between using queries provided by tracking the original query, independent queries provided in a supervised manner, and the combination of both.

\begin{figure}[t]
\scalebox{.95}{\includegraphics[width=\linewidth]{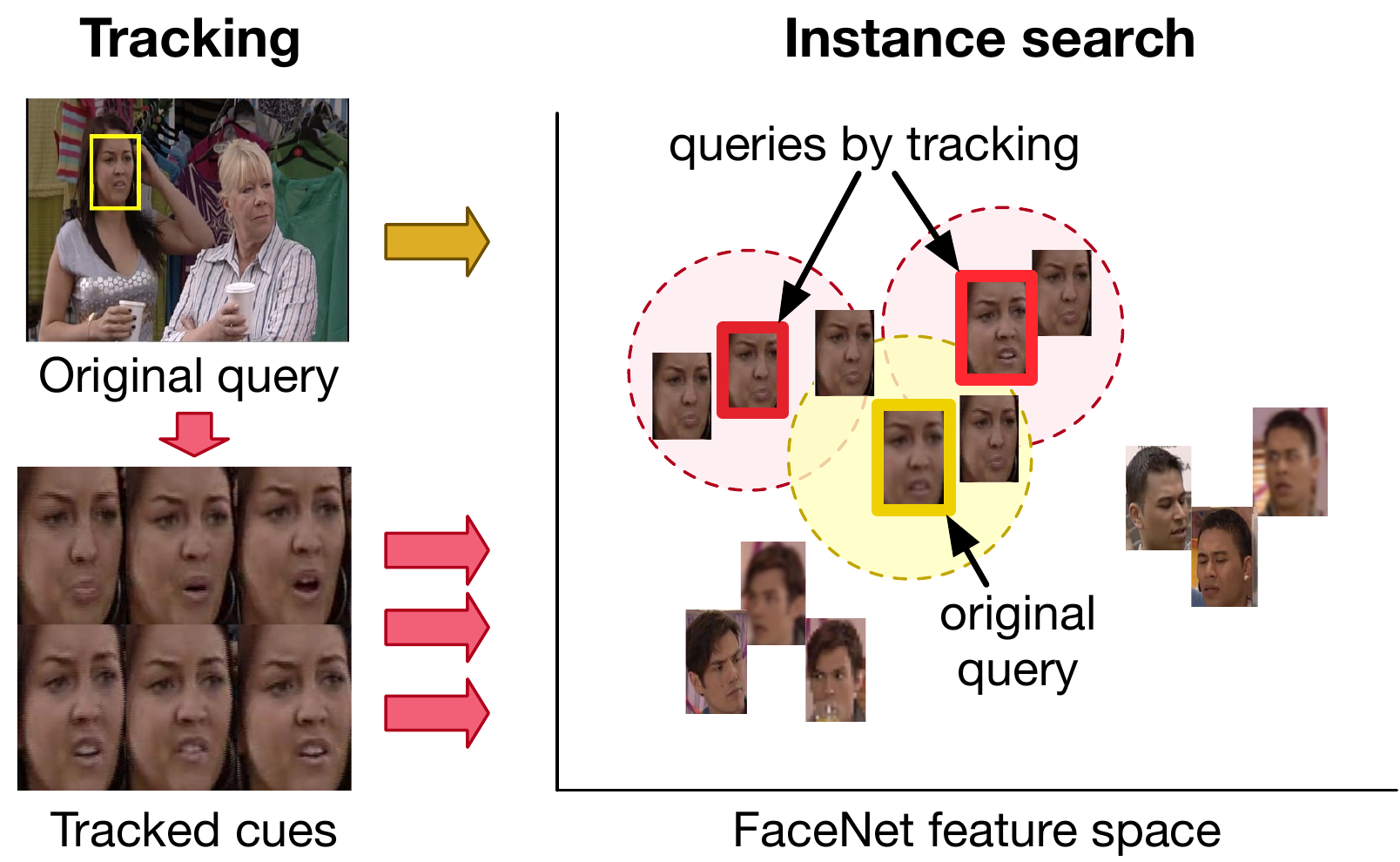}}
\caption{Single query being expanded by means of tracking. The new query examples coming from the tracked cues introduce new information of the queried person.}
\label{fig:figure1}
\end{figure}
\label{sec:intro}

\section{RELATED WORK}
\documentclass[Tracked_INS.tex]{subfiles}
\externaldocument{sq_vs_mq_plots.tex}

\textbf{Person search:}

Most of the literature on person instance search relies on CNN models to extract face features in order to look for instances corresponding to the original query. \cite{trec16_dcu,trec16_iti-certh,trec16_nii,trec16_pku,trec16_siat,trec16_whu} use VGG-faces \cite{parkhi2015VGGfaces} for face feature extraction, \cite{trec16_irim} maps the face features on a FaceNet \cite{schroff2015facenet} embedding, and \cite{trec16_chemnitz} uses a Faster-RCNN \cite{renNIPS15fasterrcnn} approach. 
\\
\\
\textbf{Multiple queries - Tracking:}

Multiple queries are shown to be useful in \cite{arandjelovic2012multiple} and \cite{zhu2014multi}. Many query expansion techniques have been tested for the instance search task. \cite{trec16_dcu} uses the first 20 ranks outputted from a first run for query expansion, \cite{trec16_pku} makes use of a first run too (top 50 ranks) to fine tune a VGG-Face CNN model and use these feature vectors to do the final search. On a more supervised way, \cite{trec16_chemnitz} annotated every instance of the main characters in an episode of a TV show to train a Faster-RCNN, while \cite{trec16_whu} looked for faces of the actors on the Internet.

In what tracking concerns, \cite{trec16_irim} and \cite{trec16_siat} make use of person tracking for query expansion, and \cite{trec16_pku} uses person tracking for person re-identification. \cite{trec16_irim} makes use of backward and forward face tracking to provide an average over a FaceNet embedding, and \cite{trec16_siat} tracks both the original query (for query expansion) and the dataset (for correlating object apparitions). 

This work studies the behavior of an instance search system when applying tracking to the given query/queries as a generic way to improve the performance of the system for free, i.e., without requiring any extra data besides the video and the query. Our proposal is to treat every tracking proposal as an independent query, and vote among the results having into account the tracking on the initial query. We study the implications of applying tracking to the instance search problem, and propose several approaches.
\label{sec:soa}

\section{instance search}
\documentclass[Tracked_INS.tex]{subfile}

We developed a person retrieval instance search system following an approach based on \cite{Hinami2016} and \cite{hinami2017}. It extracts region-based CNN features \cite{girshick2014rich} from object candidates to generate the database. They are indexed using product quantization (PQ) and an inverted index \cite{jegou2011product}, which enables to search relevant objects efficiently.

Here, we will differentiate between face feature space generation and querying.

On database generation: first, we sample the dataset videos at 1fps. Then, for each frame, we extract the faces of the people in it by means of a multi-task cascaded CNN \cite{MTCNN16}. For every face, its features are extracted using FaceNet \cite{schroff2015facenet}. These features are clustered to build an inverted index, and the PQ-compressed codes are stored in them.

On querying: given an initial mask or bounding box of the person of interest, his/her face bounding box is detected and the face features are extracted. The distances between these features from the query and the cluster centroids are calculated (in this work we used the Euclidean distance). Then, the top-k closest centroids to the query are chosen, and every face contained in each cluster is compared against the original face query to provide a similarity score to, finally, propose a ranked list of frames.

The above methodology is specified for a single, independent query. To combine multiple queries we refer to our voting scheme in Section \ref{sub:voting}, on a set of independent queries. In short, we combine the resulting ranked list from every query and do a re-ranking based on the final score of each shot. As can be seen in table \ref{tab:notrack_vs_track} (No tracking column), using multiple queries of the same person instead of a single one improves, by a huge margin, the results. On average, the difference between using a single query vs multiple queries per person is, for 2 provided queries: $\Delta mAP=+0.098$, and for 4 provided queries: $\Delta mAP=+0.200$. 

\begin{figure*}[t]
\centering
{\includegraphics[width=\linewidth]{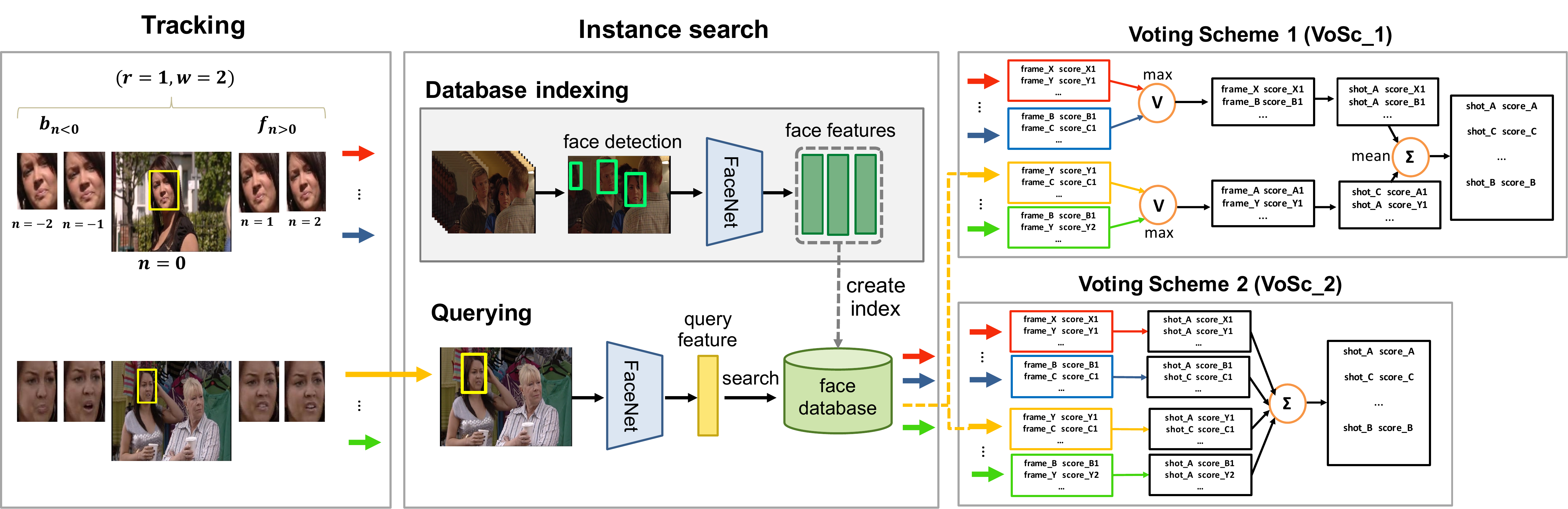}}
\caption{Our pipeline. First we track the original query example along a defined window, which will propose a set of new query examples. Then the instance search is performed individually for all the proposed query examples (original query example and query examples provided by tracking). Finally, we combine the results of every query example in order to have a ranked list of shots that contain the queried person.}
\label{fig:pipeline}
\end{figure*}
\label{sec:instance_search}

\section{TRACKED instance search}
\documentclass[Tracked_INS.tex]{subfile}

In the previous section we stated that multiple queries of the same instance provide, if the multiple queries do not contain a bad query that maximizes a bad retrieval score, a boost in performance. 

We want to extend this idea to unsupervised queries, this is, to generate query examples derived from an initial query example from a video. We achieve this by tracking, backward and forward, the original query example. Tracking will provide a new set of occurrences that will correspond to the original query, and thus, diversity among the results from the instance search system. Figure \ref{fig:pipeline} shows our pipeline.

\subsection{Tracking}
\label{sub:tracking}

The aim of tracking is to provide diversity to the given query making use of the video information. This is, given an initial query example from a video, track it forward and backward in order to provide more examples of it. For a general case, tracking would come in handy for many obvious reasons, which can be summarized in {\it automated query expansion}. Our tracker uses \cite{MTCNN16} to detect faces in a frame and provide their alignment points (eyes, nose and mouth), and \cite{wen2016discriminative} to extract the face features.  

We first position at the frame of the video corresponding to the given example $q_{n=0}$ ($n \in \mathbb{N}$), and define a temporal window $w \geq{0}$ centered over it ($w=0$ means no tracking). The backward cue $b_{n<0} \geq{0} \in \mathbb{N}$ and the forward cue $f_{n>0} \geq{0} \in \mathbb{N}$ contain the neighboring frames of the given example, where $n \in [-w,\ldots,w]$. 

We compare the distance between the face feature vector from $q_{n=0}$, $v_{n=0}$, against the feature vector of every face in the frames included in $w$, $v^{(k)}_{n}$, where $k \in \mathbb{N}$ corresponds to each face in the frame (as there can be $0 \ldots K$ faces in a frame). To do so we use the Euclidean distance. Then, the most similar face per frame is chosen and thresholded, so we get 1 or 0 examples for that frame. The resulting examples for a given query will result in $q \leq 2 \dot{w}+1$ ($2\dot{w}$ for examples provided by tracking + 1 given example). 

It is important to state that the sample rate $r$ over $n$ is not irrelevant. Let us define a sample rate $r>0 \in \mathbb{N}$. $n$ will be sampled as $n \in r \dot{[-w,\ldots,w]}$, e.g. if $n$ is sampled on a rate $r=1$ over a window $w=2$ then $n=[-2,1,0,1,2]$. This will have impact on the variance of the proposed samples by the tracked examples $q_{n \neq 0}$ with respect to the original query $q_{n=0}$. If $r=1$ the neighboring frames will be the immediate anterior and posterior frames, but if $r=20$ the neighboring frames will be further away from the given example. This means that the variance of the samples provided by tracking, in a general case, will depend on the sampling rate $r$, e.g. $r=1$ might produce little variance over the original given example while $r=20$ might produce almost independent queries.  

Caveat: the tracking is performed on the shot where the given example is provided. This means that, working in a high rate / big window configuration (e.g. $(r=20,w=5)$), some frames may fall outside the shot and, therefore, not taken in account. Further research could track over a temporal window without shot time constraint, taking in account the whole video.

\subsection{Instance search}

The instance search system is described in Section \ref{sec:instance_search}. Examples provided by the tracking step are processed independently to produce a set of ranked results (one ranked list per example).

\subsection{Voting}
\label{sub:voting}

Finally, a voting step is proposed in order to combine the results provided by the instance search system. To do so, we have worked on two possible voting schemes, $VoSc\_1$ and $VoSc\_2$. The first one assumes that the visual examples coming from the tracked cues of a provided example are dependent between them, and the second that they are independent. The original given examples are considered always as independent between them.

The goal of video instance search is to find videos from a dataset where a certain object instance appears. In the TRECVID challenge the dataset is generated from a TV show, and these videos are shots from different chapters. Our instance search system returns a scored set of frames where the queried person is likely to appear, while each one of these frames correspond to a certain shot in the dataset. The final result is an ordered list of shots depending on their score. 

Let us define $s_{n,k}^{(i)}$ as the score of containing the person to be searched for a frame $i$ in the resulting ranked list $n$ corresponding to the tracked cue of a provided example $k$, where $n \in [-w,\ldots,w]$, being $b_{q} \in [-w,\ldots,0)$ and $f_{q} \in (0,\ldots,w]$ the tracked cues defined in Section \ref{sub:tracking}, and $0$ the original query. Each of these scores are associated to a single frame in the database without repetition. 

Every visual example (original and coming from tracking) generates its own ranked list of frames likely to contain an instance of the query. $VoSc\_1$ assumes dependency between results coming from the same tracked cues, so they are merged to a single ranked list of frames by doing a max pooling (taking the maximum score for every frame). Let $s_{k}^{(i)}$ be the resulting score list of the combination between $s_{n,k}^{(i)}$ for $n \in [-w,\ldots,w]$. Then:

\begin{equation}
\label{eq:voting_1}
s_{k}^{(i)} = max(s_{-w,k}^{(i)}, \cdots ,s_{w,k}^{(i)}) 
\end{equation}

After doing this for every given example $k$, we then proceed to the combination among them. First, we map every resulting frame $i$ into its corresponding shot $u$, where $u \leq i$. Then, we combine the shots depending on their score and number of instances retrieved. This combination will result in $s^{(u)}$, which is the resulting score of the shot $u$. This combination between shots is performed as the sample mean.

\begin{equation}
\label{eq:voting_2}
s^{(u)} = \frac{1}{K} \sum_1^K (s_{1}^{(u)}, \cdots ,s_{k}^{(u)}) 
\end{equation}

$VoSc\_1$ follows this two-step voting scheme (merge the results from tracked cues and combine them with other given examples if any), and $VoSc\_2$ only follows the second step (combine all the proposals as if they were all given examples). Here, a discussion on how we consider a query to be independent or not arises. Generally speaking, a query coming from the tracking phase could be considered as independent when it differs largely (in terms of pose, illumination...) within its neighbors. We expect an independent query to produce different results than the queries that are only a little variation of the original query, which should have similar results. In this case, the results in Figure \ref{fig:mxmn_vs_mnmn} conclude that the tracked cues should not be considered as independent.
\label{sec:tracked_instance_search}

\section{EXPERIMENTS}
\documentclass[Tracked_INS.tex]{subfile}

{\bf Dataset}

We used the dataset provided by TRECVID for instance search task. It is composed by 244 chapters of the BBC show {\it Eastenders}, resulting in 464 hours of video data separated in 471527 shots.  

\subsection{Ground truth generation}
\vspace{-1ex}
In order to evaluate this work, we generated the ground truth for person retrieval (independent of place) on the TRECVID \cite{2016trecvidawad} 2016 instance search task. To do so, we proceeded in the following manner: for every resulting ranked list of shots in every experiment (e.g. $(r=2,w=3)$), we evaluated the first 300 results. If the queried person appeared in the resulting shot, we added that shot to the ground truth for the corresponding query. We evaluated the first 300 results for all the combinations between window size (0 to 5) and rate (1,2,5,20). We ended up with a ground truth of 1143 samples per person on average (8006 samples in total), being {\it Stacey} (query 9174) the query with maximum number of annotated shots (1583), and {\it Jim} (query 9162) the query with minimum number of annotated shots (409) (see TRECVID INS 2016 task for details). This ground truth will be made publicly available to help further research.

\subsection{Tracked instance search}

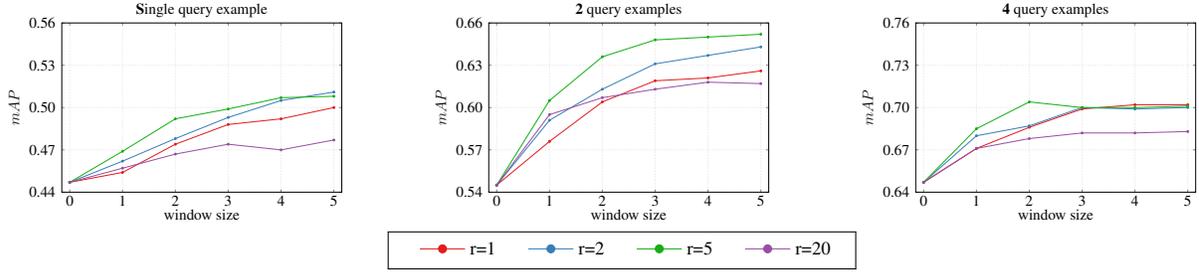
\begin{figure*}[t!]
\centering
\pgfplotstableread{data/sqt_vs_mqt_sq_mAP.txt}\datasqmAP
\pgfplotstableread{data/sqt_vs_mqt_mq2_mAP.txt}\datamqtwomAP
\pgfplotstableread{data/sqt_vs_mqt_mq4_mAP.txt}\datamqfourmAP
\hspace{5px}
\resizebox{.25\linewidth}{!}{\begin{tikzpicture}[/pgfplots/width=\linewidth, /pgfplots/height=0.64\linewidth, /pgfplots/legend pos=south east]
    \begin{axis}[
    	ymin=0.44, ymax=0.56,
        xmin=0.0, xmax=5.0,
        ylabel=$mAP$,
        xlabel=window size,
		font=\Huge,
        grid=both,
		grid style=dotted,
        xlabel shift={-2pt},
        ylabel shift={10pt},
        xmode=linear,
        yticklabel style = {xshift=-0.25cm},
        xticklabel style = {yshift=-0.1cm},
        ytick={0.44,0.47,0.50,0.53,0.56},
		yticklabels={0.44,0.47,0.50,0.53,0.56},
        xtick=data,
	    xticklabels from table={\datasqmAP}{w},
        enlarge x limits=0.03,
        title={\textbf Single query example},
        ]

		\addplot+[color=r1,mark=*,mark size=1.5,mark options={fill=r1}] table[x expr=\coordindex,y=r1]{\datasqmAP};
        \label{fig:sq_vs_mq_nt:1}

        \addplot+[color=r2,mark=*,mark size=1.5, mark options={fill=r2}] table[x expr=\coordindex,y=r2]{\datasqmAP};
        \label{fig:sq_vs_mq_nt:2}
        
        \addplot+[color=r5,mark=*,mark size=1.5,mark options={fill=r5}] table[x expr=\coordindex,y=r5]{\datasqmAP};
        \label{fig:sq_vs_mq_nt:3}
        
        \addplot+[color=r20,mark=*,mark size=1.5,mark options={fill=r20}] table[x expr=\coordindex,y=r20]{\datasqmAP};
        \label{fig:sq_vs_mq_nt:4}
                          
    \end{axis}
\end{tikzpicture}}
\hspace{0.06\linewidth}
\resizebox{.25\linewidth}{!}{\begin{tikzpicture}[/pgfplots/width=\linewidth, /pgfplots/height=0.64\linewidth, /pgfplots/legend pos=south east]
    \begin{axis}[
    	ymin=0.54, ymax=0.66,
        xmin=0.0, xmax=5.0,
        ylabel=$mAP$,
        xlabel=window size,
		font=\Huge,
        grid=both,
		grid style=dotted,
        xlabel shift={-2pt},
        ylabel shift={10pt},
        xmode=linear,
        yticklabel style = {xshift=-0.25cm},
        xticklabel style = {yshift=-0.1cm},
        ytick={0.54,0.57,0.60,0.63,0.66},
		yticklabels={0.54,0.57,0.60,0.63,0.66},
        xtick=data,
	    xticklabels from table={\datamqtwomAP}{w},
        enlarge x limits=0.03,
        title={\textbf 2 query examples}     
        ]

		\addplot+[color=r1,mark=*,mark size=1.5,mark options={fill=r1}] table[x expr=\coordindex,y=r1]{\datamqtwomAP};
        \label{fig:sq_vs_mq_nt:1}
        
        \addplot+[color=r2,mark=*,mark size=1.5, mark options={fill=r2}] table[x expr=\coordindex,y=r2]{\datamqtwomAP};
        \label{fig:sq_vs_mq_nt:2}
        
        \addplot+[color=r5,mark=*,mark size=1.5,mark options={fill=r5}] table[x expr=\coordindex,y=r5]{\datamqtwomAP};
        \label{fig:sq_vs_mq_nt:3}
        
        \addplot+[color=r20,mark=*,mark size=1.5,mark options={fill=r20}] table[x expr=\coordindex,y=r20]{\datamqtwomAP};
        \label{fig:sq_vs_mq_nt:4}
                          
    \end{axis}
\end{tikzpicture}}
\hspace{0.06\linewidth}
\resizebox{.25\linewidth}{!}{\begin{tikzpicture}[/pgfplots/width=1\linewidth, /pgfplots/height=0.64\linewidth, /pgfplots/legend pos=south east]
    \begin{axis}[
    	ymin=0.64, ymax=0.76,
        xmin=0.0, xmax=5.0,
        ylabel=$mAP$,
        xlabel=window size,
		font=\Huge,
        grid=both,
		grid style=dotted,
        xlabel shift={-2pt},
        ylabel shift={10pt},
        xmode=linear,
        yticklabel style = {xshift=-0.25cm},
        xticklabel style = {yshift=-0.1cm},
        ytick={0.64,0.67,0.70,0.73,0.76},
		yticklabels={0.64,0.67,0.70,0.73,0.76},
        xtick=data,
	    xticklabels from table={\datamqfourmAP}{w},
        enlarge x limits=0.03,
        title={\textbf 4 query examples},
        ]

		\addplot+[color=r1,mark=*,mark size=1.5,mark options={fill=r1}] table[x expr=\coordindex,y=r1]{\datamqfourmAP};
        \label{fig:sq_vs_mq_nt:1}
        
        \addplot+[color=r2,mark=*,mark size=1.5, mark options={fill=r2}] table[x expr=\coordindex,y=r2]{\datamqfourmAP};
        \label{fig:sq_vs_mq_nt:2}
        
        \addplot+[color=r5,mark=*,mark size=1.5,mark options={fill=r5}] table[x expr=\coordindex,y=r5]{\datamqfourmAP};
        \label{fig:sq_vs_mq_nt:3}
        
        \addplot+[color=r20,mark=*,mark size=1.5,mark options={fill=r20}] table[x expr=\coordindex,y=r20]{\datamqfourmAP};
        \label{fig:sq_vs_mq_nt:4}
                          
    \end{axis}
\end{tikzpicture}}

\vspace{1mm}
\hspace{8mm}
\framebox[0.35\linewidth]{
\begin{minipage}{\textwidth}
\centering
\scriptsize
\begin{tabular}{llll}
\ref{fig:sq_vs_mq_nt:1} r=1 & \ref{fig:sq_vs_mq_nt:2} r=2 & \ref{fig:sq_vs_mq_nt:3} r=5 & \ref{fig:sq_vs_mq_nt:4} r=20
\end{tabular}
\end{minipage}}
\caption{$mAP$ for different window sizes $(w=0,1,2,3,4,5)$, where $w=0$ corresponds to no tracking, and rates $(r=1,2,5,20)$ for a single given query example (left), 2 provided query examples (middle), and 4 provided query examples (right), using voting scheme $VoSc\_1$.}
\label{fig:sqt_vs_mqt}
\end{figure*}

The combination of tracking and voting make the system performance improve. As seen in table \ref{tab:notrack_vs_track}, by applying our method with an $(r=2,w=5)$ configuration, the $mAP$ has a percentage increase of $14.3\%$ for single query tracking. For multiple queries, the system improves by $18\%$ when having 2 provided queries, and by $8.2\%$ for 4 provided queries. In figure \ref{fig:sqt_vs_mqt} different configurations of our method are studied. 

Best configuration for each case scenario makes $mAP$ increase by: $\Delta mAP=+0.064$ for a single query $(r=2,w=5)$, $\Delta mAP=+0.108$ for 2 queries $(r=5,w=5)$, and $\Delta mAP=+0.057$ for 4 queries $(r=5,w=2)$.

A discussion on different frame rates arises. We stated along the work that a higher rate should lead to better results due to a larger variation of the produced examples by tracking. In figure \ref{fig:sqt_vs_mqt}, rate $r=5$ outperforms $r=1$ and $r=2$ but, clearly, a sampling rate of $r=20$ performs way worse than the others. This has two big caveats: first, it is more likely to find the same person's face in closer frames. As we work on a TV show dataset, there are many scene changes where the original query might not be present. If a sampled frame does not contain an instance of the original query (the person to track is not present), we are not able to use that as extra information for the instance search. 

Second caveat: as stated in section \ref{sub:tracking}, we work on a shot-defined length per query. This means that, exclusively, we track inside the specific shot containing the original query. The reasoning is that we can only be sure that instances of the original query will be present inside that single shot. 
Further research could explore the possibility on tracking along the video (intra search) until finding a specified number of instances of the original query for the further instance search (inter search). 

\begin{table}[h]
\centering
\scalebox{1}{\begin{tabular}{c|cc}
Examples given & No tracking & Tracking \\
\hline \\[-2mm]
1 & 0.447       & {\bf 0.511}   \\
2 & 0.545       & {\bf 0.643}    \\
4 & 0.647       & {\bf 0.700}   
\end{tabular}}
\caption{Average $mAP$ with and without tracking for the 4 examples per query provided by TRECVID (2016) INS task. The configuration used is $(r=2,w=5)$ for tracking and $VoSc\_1$ for voting.}
\label{tab:notrack_vs_track}
\end{table}

\subsection{Voting}
In Figure \ref{fig:mxmn_vs_mnmn} we compare the resulting $mAP$ considering the cues coming from tracking dependent or independent, this is, merging them following the defined two-step voting scheme ({\it VoSc\_1}) or directly mapping each frame into its corresponding shot and combining them by applying directly the second step ({\it VoSc\_2}). Results show that by merging first the results coming from the tracked cues better results are achieved, rather than directly combining the outputs from the instance search for all visual examples (tracked and provided) as they were independent among them. 

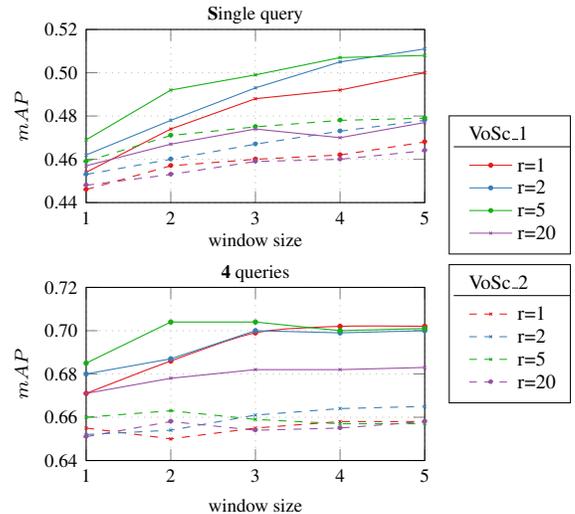
\begin{figure}[h!]
\centering
\scalebox{.95}{
\begin{minipage}{0.7\linewidth}
\resizebox{\linewidth}{!}{\begin{tikzpicture}[/pgfplots/width=\linewidth, /pgfplots/height=0.64\linewidth, /pgfplots/legend pos=south east]
    \begin{axis}[
    	ymin=0.44, ymax=0.52,
        xmin=1.0, xmax=5.0,
        ylabel=$mAP$,
        xlabel=window size,
        font=\scriptsize,
        grid=both,
		grid style=dotted,
        xlabel shift={-1mm},
        ylabel shift={0pt},
        xmode=linear,
        ytick={0.44,0.46,0.48,0.50,0.52},
		yticklabels={0.44,0.46,0.48,0.50,0.52},
        xtick=data,
	    xticklabels={1,2,3,4,5},
        enlarge x limits=0.00,
		title style={yshift=-2.5mm},
        title={\textbf Single query},
        ]

		\addplot[color=r1,dashed,mark=*, mark size=0.8,mark options={fill=r1}] coordinates{(1,0.446)(2,0.457)(3,0.460)(4,0.462)(5,0.468)};
        \label{fig:mnmn:r1}
        \addplot[color=r1,mark=x, mark size=0.8,mark options={fill=r1}] coordinates{(1,0.454)(2,0.474)(3,0.488)(4,0.492)(5,0.50)};
        \label{fig:mxmn:r1}
     
        \addplot[color=r2,dashed,mark=*, mark size=0.8,mark options={fill=r2}] coordinates{(1,0.453)(2,0.460)(3,0.467)(4,0.473)(5,0.478)};
        \label{fig:mnmn:r2}
        \addplot[color=r2,mark=x, mark size=0.8,mark options={fill=r2}] coordinates{(1,0.462)(2,0.478)(3,0.493)(4,0.505)(5,0.511)};
        \label{fig:mxmn:r2}
                      
        \addplot[color=r5,dashed,mark=*, mark size=0.8,mark options={fill=r5}] coordinates{(1,0.459)(2,0.471)(3,0.475)(4,0.478)(5,0.479)};
        \label{fig:mnmn:r5}
        \addplot[color=r5,mark=x, mark size=0.8,mark options={fill=r5}] coordinates{(1,0.469)(2,0.492)(3,0.499)(4,0.507)(5,0.508)};
        \label{fig:mxmn:r5}
                
        \addplot[color=r20,dashed,mark=*, mark size=0.8,mark options={fill=r20}] coordinates{(1,0.448)(2,0.453)(3,0.459)(4,0.460)(5,0.464)};
        \label{fig:mnmn:r20}
        \addplot[color=r20,mark=x, mark size=0.8,mark options={fill=r20}] coordinates{(1,0.457)(2,0.467)(3,0.474)(4,0.470)(5,0.477)};
        \label{fig:mxmn:r20}
    \end{axis}
\end{tikzpicture}}

\resizebox{\linewidth}{!}{\begin{tikzpicture}[/pgfplots/width=\linewidth, /pgfplots/height=0.64\linewidth, /pgfplots/legend pos=south east]
    \begin{axis}[
    	ymin=0.64, ymax=0.72,
        xmin=1.0, xmax=5.0,
        ylabel=$mAP$,
        xlabel=window size,
        font=\scriptsize,
        grid=both,
		grid style=dotted,
        xlabel shift={0pt},
        ylabel shift={0pt},
        xmode=linear,
        ytick={0.64,0.66,0.68,0.70,0.72},
		yticklabels={0.64,0.66,0.68,0.70,0.72},
        xtick=data,
	    xticklabels={1,2,3,4,5},
        enlarge x limits=0.00,
		title style={yshift=-2.5mm},
        title={\textbf 4 queries},
        ]

		\addplot[color=r1,dashed,mark=x, mark size=0.8,mark options={fill=r1}] coordinates{(1,0.655)(2,0.650)(3,0.655)(4,0.658)(5,0.658)};
        \label{fig:mnmn:r1}                 
        \addplot[color=r1,smooth,mark=*, mark size=0.8,mark options={fill=r1}] coordinates{(1,0.671)(2,0.686)(3,0.699)(4,0.702)(5,0.702)};
        \label{fig:mxmn:r1}
     
        \addplot[color=r2,dashed,mark=x, mark size=0.8,mark options={fill=r2}] coordinates{(1,0.652)(2,0.654)(3,0.661)(4,0.664)(5,0.665)};
        \label{fig:mnmn:r2}        
        \addplot[color=r2,mark=*, mark size=0.8,mark options={fill=r2}] coordinates{(1,0.680)(2,0.687)(3,0.700)(4,0.699)(5,0.700)};
        \label{fig:mxmn:r2}
                    
        \addplot[color=r5,dashed,mark=x, mark size=0.8,mark options={fill=r5}] coordinates{(1,0.660)(2,0.663)(3,0.659)(4,0.657)(5,0.657)};
        \label{fig:mnmn:r5}
        \addplot[color=r5,mark=*, mark size=0.8,mark options={fill=r5}] coordinates{(1,0.685)(2,0.704)(3,0.704)(4,0.700)(5,0.701)};
        \label{fig:mxmn:r5} 
        
        \addplot[color=r20,dashed,mark=*, mark size=0.8,mark options={fill=r20}] coordinates{(1,0.651)(2,0.658)(3,0.654)(4,0.655)(5,0.658)};
        \label{fig:mnmn:r20}
        \addplot[color=r20,mark=x, mark size=0.8,mark options={fill=r20}] coordinates{(1,0.671)(2,0.678)(3,0.682)(4,0.682)(5,0.683)};
        \label{fig:mxmn:r20}

    \end{axis}
\end{tikzpicture}}
\end{minipage}
\hspace{0mm}
\begin{minipage}{0.15\linewidth}
\centering
\footnotesize
\framebox[\width]{
\begin{tabular}{l}
VoSc\_1 \\ \hline \\[-2mm] 
\ref{fig:mxmn:r1} r=1 \\ 
\ref{fig:mxmn:r2} r=2 \\ 
\ref{fig:mxmn:r5} r=5 \\
\ref{fig:mxmn:r20} r=20 \\
\end{tabular}}
\\[1mm]
\framebox[\width]{
\begin{tabular}{l}
VoSc\_2 \\ \hline \\[-2mm] 
\ref{fig:mnmn:r1} r=1 \\ 
\ref{fig:mnmn:r2} r=2 \\ 
\ref{fig:mnmn:r5} r=5 \\
\ref{fig:mnmn:r20} r=20 \\
\end{tabular}}
\end{minipage}}
\caption{mAP for a single query example and multiple query examples (4) for two different voting schemes (VoSc\_1 and VoSc\_2). VoSc\_1 considers tracked cues as dependent and provided queries as independent, VoSc\_2 considers both tracked and given query examples as independent.}
\label{fig:mxmn_vs_mnmn}
\end{figure}

\subsection{On the amount of extra data}

The extra data feeded into the system by means of tracking helps to have a relative gain of almost a $15 \%$ (1 provided query example) for free (using only the given query and the associated video). It is interesting to note in figure \ref{fig:sqt_vs_mqt} that the system seems to get to a saturation point (for 2 and 4 provided queries), where increasing the amount of data provided by the tracker does not improve the performance at any sampling rate. Further research could explore the performance of the system when using larger windows on a single query case, as it seems that the saturation in that case has not yet been reached, to test if the system could perform as well as having multiple independent queries. 
\label{sec:experiments}

\section{Conclusions and further research}
\documentclass[Tracked_INS.tex]{subfile}

We have investigated the performance of incorporating tracking to the instance search task, and a way to combine the resulting scored lists. Having a query and its associated video, we can state that using tracking as a generic query expansion the performance of the system can be improved for free (without requiring any more data). We have also observed that multiple independent queries provide a considerable increase of performance. Further research could aim at providing more variable query examples of the same person, so they could be considered as independent queries.
\label{sec:conclusions}

\bibliographystyle{IEEEbib}
\bibliography{refs}

\end{document}